# Comment: The 2005 Neyman Lecture: Dynamic Indeterminism in Science

**Grace L. Yang**

*Key words and phrases:* Neyman, survey sampling, survival analysis, martingales, effect of radiation, shoulder effect, competing risks.

Professor Brillinger wrote a very stimulating paper on Neyman's life history and some of his contributions to applied statistics. The paper's central theme is to review how Neyman used stochastic processes in data analysis. The paper contains a number of illuminating examples of Neyman and of Brillinger with other collaborators. I am honored to have been invited to be a discussant.

Professor Brillinger quoted Neyman (1960), "The time has arrived for the theory of stochastic processes to become an item of usual equipment of every applied statistician." In the post-Neyman era, data come in our way fast and in all forms, such as streams, functions, manifolds, random shapes, trees and images. The importance of the theory of stochastic processes in applied statistics cannot be overemphasized.

Brillinger's observation of Neyman's thought processes in conducting applied research resonates with me. My discussion will be primarily to amplify it from a somewhat different perspective, namely from Neyman's teaching and his research projects on sampling and cancer. Included in the discussion will be recalls of some of my personal experience having Neyman as a teacher. Neyman's sampling and cancer projects are selected in this discussion in part because of their broad impact which appears to be not a focus of Brillinger's paper. Although Neyman's sampling work does not involve stochastic processes, it fits the title of Brillinger's paper "Dynamic Indeterminism in Science." Neyman had engaged in cancer research for many years until his death in 1981. His cancer research (including survival analysis) used Markov processes extensively. Neyman's contribution to survival analysis links nicely to Brillinger's view on the importance of point processes. Special attention will be paid to Neyman's *Lecture Notes and Conferences on Mathematical Statistics* (1938, 1952) in which Neyman introduced many fundamental statistical concepts and statistical theory, and discussed his views on statistical research which I believe are still very current.

## 1. NEYMAN AS A TEACHER AND HIS PROBLEM-DRIVEN APPROACH

I was a student in several of Neyman's classes and a regular in his weekly seminar. My thesis advisor, Lucien Le Cam, sent me to Neyman's classes. Actually, Neyman and Le Cam were like co-advisors to many Ph.D. students of theirs. Neyman would say, "Go ask Mr. Le Cam" or the other way around.

Neyman did not use notes and the lectures were based mostly on his research work. A typical lecture started with a description of a physical problem which was then followed by a discussion of the chance mechanisms operating in the physical phenomenon, and the construction of a model for the data. Next he would pose a statistical hypothesis for testing or developing some estimation procedures. We learned firsthand why he introduced such statistical concepts and methods. Neyman's way of first studying a physical problem and leading to the eventual development of a statistical procedure is quite opposite to the practice of starting with some available statistical methods and applying them to a physical problem. The order of attacking a scientific problem seems reversed.

*Grace L. Yang is Program Director for Statistics Program, Division of Mathematical Sciences, National Science Foundation, 4201 Wilson Blvd, Arlington, Virginia 22230, USA (e-mail: gyang@nsf.gov) and Professor, Department of Mathematics, University of Maryland, College Park, Maryland.*

In these classes, we went through stochastic processes and solved differential equations for probability generating functions with a wide range of applications. For a while we had seminar every Wednesday evening, discussing models of carcinogens and passing around photos of tumors of all shapes (not pretty). Students were called to the blackboard for questions and discussions. Sometimes, the seminars could last until 11 PM and Neyman would take us to Shattuck Avenue for cake and ice cream afterward. Neyman cared a great deal about his students. Once a student did not show up in his class for a couple of weeks. Neyman was worried and knocked on the student's apartment but there was no answer. He had a policeman break into the apartment. It turned out the student had taken a trip without informing Neyman. He was a tremendous mentor and continued to provide valuable advice to his formal students throughout his life. What a privilege I had. I had opportunities of seeing Neyman when he came to Washington for meetings or to "shop for money" as he put it. When in Washington, he stayed at the Cosmos Club. I recall, when we went to see him, while his other academic siblings (Bob Traxler and Tom Darden) could enter the club through the front door, I could only use the side door entrance. That was in the 1970s. Women were reminded often of their lack of social status.

Many noted that Neyman had a great appreciation of Lebesgue's theory of integration. Indeed, Neyman liked to ask us, "Do you know there is a difference between the improper Riemann integral and the Lebesgue integral?" Cloud seeding was one of Neyman's long-term projects. Randomization of the decision to seed or not seed was strictly observed. He had assistants in his laboratory flip coins to decide on seeding or not seeding in his experiments in Europe. When talking about competing risks, he would ask if we have seen a death certificate. Le Cam (a student of Neyman) (1995) describes Neyman, "He was always full of energy and ideas and 'imprinted' them on his students in courses or in individual contacts."

A recent book by Calvin Moore (2007) on the history of the Berkeley Mathematics Department gives a vivid account of Neyman's early days in the Department of Mathematics, and his 17 years of struggle to form the Department of Statistics. "Neyman continued to agitate for an independent department of statistics (Moore, 2007, page 83)." The Department of Statistics was established in 1955. Neyman would not give up.

## 2. NEYMAN'S TRIUMPHANT 1937 U.S. TOUR AND ADOPTION OF SAMPLING IN U.S. 1940 CENSUS

At the invitation of Edward Deming, Neyman toured the United States in the spring of 1937 for the first time and gave lectures at the Graduate School of the U.S. Department of Agriculture. His lectures were published in *Lecture Notes and Conferences on Mathematical Statistics, 1938.* The second edition *Lecture Notes and Conferences on Mathematical Statistics and Probability* was published in 1952. Notice the addition of Probability in the title. The second edition differs substantially from the first edition because, according to Neyman, of the extraordinary development of the economics and stochastic processes (Doob and Feller's work on stochastic processes). Thus at least since the early 1950s, Neyman had used stochastic processes extensively in his applied work. Neyman's lectures at the USDA included his revolutionary paper on survey sampling (1934) which marked a new era in sampling theory. At that time, the representative method of extracting information used by A. L. Bowley (1913) became very popular among statisticians in different countries. The popularity was partly due to the scarcity of resources and shortness of time for an exhaustive research. There are two aspects of the representative method. One of them is called the method of random sampling and the other the method of purposive selection. According to Neyman, the two kinds of methods were discussed by A. L. Bowley (1925) and they are treated as it were on equal terms, as being equally to be recommended. Much the same attitude has been expressed in a ISI report (see Jensen, 1925). Twenty years later, Neyman's paper (1934) points out the logical distinction between these two methods. He cautioned the use of purposive selection whose success is rather exceptional. Neyman's paper systematically develops the theory of stratified random sampling on the basis of random sampling. The concept of confidence intervals was also introduced in this paper. Neyman's work had a significant influence on the adoption of sampling procedure in the U.S. 1940 census. See N. Mann (1994) on E. Deming. Recounted M. Hansen (1987), "Neyman's paper and the visit... contributed much to the welfare of the U.S. and to the future acceptance of sampling, at least in the Bureau of Census."

Neyman's probability sampling and optimum allocation of sample sizes have been used to this day. By now, the notion and the practice of sampling have become a way of life and inseparable from scientific investigation, be it nutrition survey, political polls, clinical trials, or sample size determination.

Neyman believed "problems of science are a breeding ground of novel mathematical disciplines." With the advent of computer technology, massive amounts of data are coming our way in every direction. The breeding ground is unprecedentedly fertile. Neyman's approach to mining survey data (massive enough) and developing a sampling theory is an excellent example of mining data albeit in cyber space.

## 3. NEYMAN, MARKOV PROCESSES AND SURVIVAL ANALYSIS

Neyman was taken by Markov processes. He used them in cancer research. The following are two examples. The second example especially points to the importance of constructing stochastic models in studying the effect of radiation.

For many years, Neyman worked on cancer research and the chance mechanism of carcinogenesis. He used his own money to fund a conference on probability models and cancer in July, 1981. He died a few weeks later on August 5, 1981. The proceedings of the conference were published posthumously in 1982 (Le Cam and Neyman, eds.). His cancer research addressed a wide range of topics including patient survival probability in clinical trials (more traditional biostatistics problems), modeling cancer growth at the cellular level, and the effects of radiation on single cells at the DNA level. I would mention two of his contributions, his work with Fix and with Puri.

(1)  Neyman–Fix Competing Risks model.

Neyman became interested in problems with evaluating the effects of breast cancer treatment discussed at a meeting in New York in 1949. Subsequently, Fix and Neyman (1951) used a four-state homogeneous Markov chain to model the status of a patient transferring between the states of recovery and relapse until she is either lost to follow-up (or censored in modern terminology) or enters the (absorbing) state of death. The paper gives a detailed discussion about the classification of states and their connections to the available observable data provided by two doctors. From the Markov model the probability of a patient surviving beyond a specified time in the presence of computing risks of relapse and censoring was estimated, and the risks (or transitions rates) of moving from one state to another were estimated. This survival probability is used to evaluate the effectiveness of a treatment method or to compare two different treatments. This applied work created a new statistical theory. The notion of competing risks and the model introduced by Neyman and Fix laid the foundation for the future development of the theory of competing risks. The extension of this work was carried out by his students, Chin-Long Chiang in life-table constructions and medical follow-up studies (1968) and A. Tsiatis (1975, communicated by Neyman to PNAS), among others. Tsiatis addressed the nonidentifiability problem of competing risks. Fix and Neyman (1951) were concerned about the validity of the assumption of constant risks in their model. An extension to time-dependent competing risks (or nonparametric analysis) can be found in a paper of B. Altshuler (1970). This paper was communicated by Neyman to the *Mathematical Biosciences*. I am unable to find any information about the circumstance under which this investigation was carried out. Was Altshuler a visitor of Neyman, of which Neyman had many? It is worthwhile to note that Altshuler (1970) is one of the earliest papers addressing the estimation of a cumulative hazard function $\Lambda(t)$. Altshuler used it to construct an estimator of the survival probability (beyond time $t$) of a subject in the presence of competing risks. His result generalizes the celebrated Kaplan–Meier estimator. The model used by Altshuler can be recast into a finite-state nonhomogeneous Markov chain with one absorbing state (death!) which was later studied in Aalen's thesis (1975, supervised by Le Cam).

The product-limit form of the Kaplan–Meier type of estimators made their analytical study challenging. A breakthrough occurred in Aalen's thesis (published in 1978) that solved some long outstanding theoretical problems regarding the optimality and properties of the Kaplan–Meier type of estimators. [A key step to Aalen's success was the formulation of the cumulative hazard function and its estimator in terms of counting processes and compensators with that, the martingale calculus applies.] It is fitting to mention here that the counting process approach was pointed out to Aalen by D. Brillinger (duly acknowledged by Aalen); a testimony to the powerful tools of stochastic differential equations in solving real life (and death) problems. The martingale method opened a new way of solving analytical problems in survival analysis whose results have populated statistical literature in the last thirty-some years.

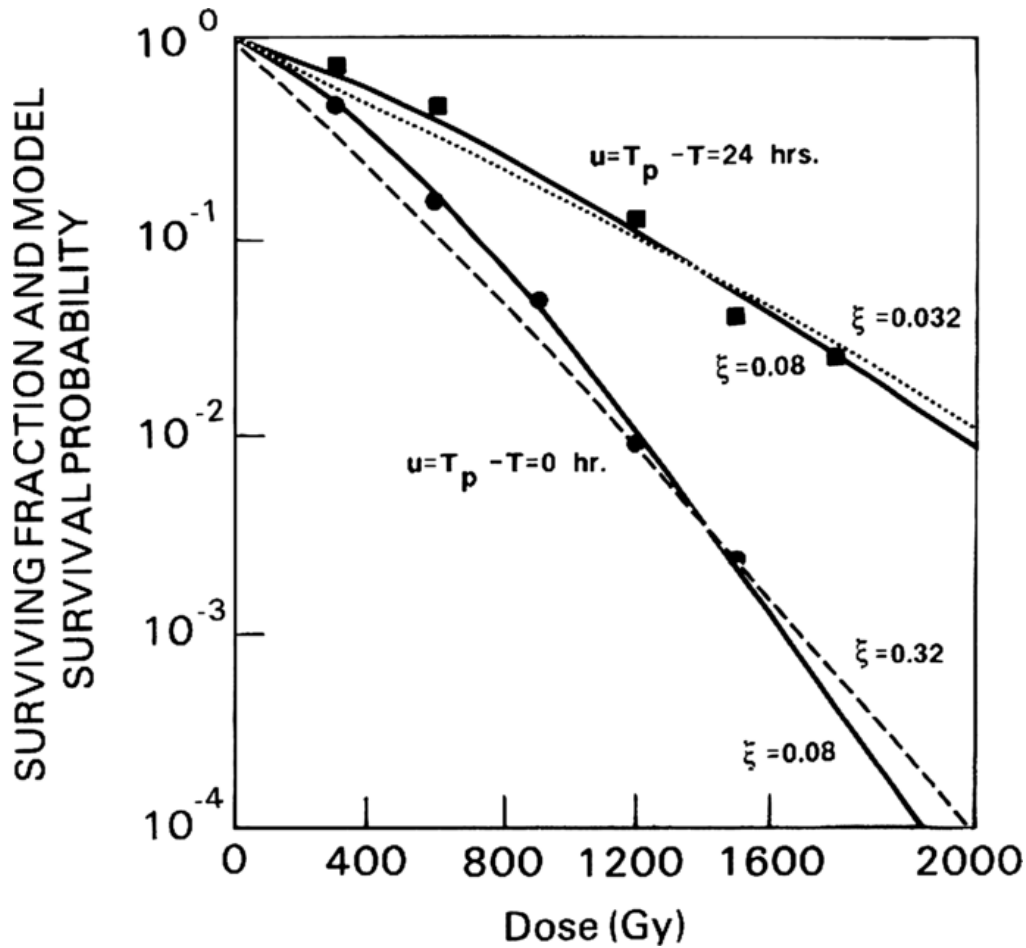

Fig. 1. *Surviving fractions of yeast cells as a function of doses irradiated with 30-MeV electrons at dose rate $r = 7800$ Gy/h. Solid lines denote the least-squares fit to the model. Dashed lines represent the effect of changing parameters $\xi = 0.32$ and $a = 1.08/h$, with all other parameters unchanged. Experimental data indicated by filled circles and filled squares from Frankenberg-Schwager et al. (1980). [First published as Figure 3 in Yang and Swenberg (1991), "Stochastic modeling of dose-response for single cells in radiation experiments," Math. Scientist, vol. 16, pages 46–65. Copyright © Applied Probability Trust 1991.]*

Markov–Branching model for effect of radiation. (2)

At the DNA level, Neyman and his students investigated the effect of ionizing radiation on single cells. The survival probability of single cells in response to dose of radiation is used as a measure of the effect. The cell mutation probability is another measure. Understanding the dose-response relationship clearly has therapeutic implications in developing criteria for either diagnosis or treatment of cancer. Moreover, radiation effects are readily observable at high doses, whereas for many matters of public policy, such as environmental cancer risk assessment and development of radioepidemiological tables for computing the probability of causation of cancer as mandated by Public Law 97–414, one needs the dose-response relationship at low doses. Low-dose experiments are very difficult to perform (if they are possible at all) and mathematical models become almost the only tool available to infer information about low-dose responses.

In radiation and biophysics literature, cell survival probability is typically modeled by $\exp(-\alpha D - \beta D^2)$ where $D$ denotes the dose, the so-called linear quadratic model [see Le Cam and Neyman (1982)]. The presence of a quadratic term is known as the shoulder effect (concave) in the log survival curve (see Figure 1). The shoulder effect is a critical experimental finding with serious implications. It implies that the radiation (low LET, such as X-rays) up to certain dose level has little effect on cell survival. The molecular mechanisms used to explain the quadratic term (the shoulder) differ significantly among researchers resulting in different models (same mathematical form but different interpretation); see Yang and Swenberg (1991) and references therein. In these models, the chance mechanism has not been systematically included by following the experimental protocol. Therefore it is difficult to sort out major experimental parameters that affect the cell survival in these models and their relations to the parameters $\alpha$ and $\beta$.

An elaborate stochastic model of a radiation experiment that considers the chance mechanisms of energy deposition, biological responses and design of the experiment was developed by Neyman and Puri (1976, 1981).

In simple terms, a radiation experiment consists of counting the proportion of cells that survive the irradiation of a given dose; the actual procedure, however, is very involved. The survival of a single cell is neither directly nor immediately observable after irradiation. The survival of a cell is thus defined by its proliferative ability to form a colony of a given (observable) size within a specified time after irradiation. Without observations, mathematical model is almost the only tool available to study the evolution of cells after irradiation.

A cell can survive radiation damage if the radiation-induced lesions are repaired completely, or survive as a mutant if it is repaired incorrectly, or be inactivated and unable to divide (death of a cell). A mutant can divide and may lead to a cancerous growth.

The Neyman–Puri model assumes the following:

1. Energy deposition. The primary radiation particles reach the cell according to a Poisson process with rate $\lambda(t)$ per unit time and unit volume.

2. Branching of primary radiation particles. Each primary radiation particle generates a random number $M$ of "spurs" with probability generating function $g(s)$. Each spur has a probability $\pi_1$ of generating a potentially lethal lesion, probability $\pi_2$ of generating an irreparable lesion (a lethal lesion) and probability $1 - \pi_1 - \pi_2$ of generating no lesion in the cell.

3. Cell's repair and misrepair mechanism. The evolution of the cell during and after radiation is modeled by a vector-valued Markov process $\{(X_t, Y_t, Z_t); t \geq 0\}$, where $X_t$ is the number of potentially lethal lesions in the cell at time $t$, $Y_t$ is the number of mutated lesions in the cell at time $t$ and $Z_t$ is the number of lethal lesions the cell has experienced up to time $t$.

Deriving the probability generating function of the process $\{(X_t, Y_t, Z_t); t \geq 0\}$ allows one to calculate the cell's survival probability and the mutation probability at any specified time.

The Armed Forces Radiation Research Laboratory (AFRRI) paid special attention to the Neyman–Puri model. I was contacted by Dr. Charles Swenberg of ARRFI which led to our collaboration to study the effects of radiation. We picked up the work left by Neyman and Puri who died in 1989. The Neyman–Puri model was given a careful examination by comparing it step by step with the protocol of the radiation experiment performed in Dr. Swenberg's laboratory. Our study resulted in modifying the Neyman–Puri formulation by including the cell repair time and nonlinear initiation of lesions. Figure 1 shows a fit of survival probability, and a fit of mutation probability is given in Figure 2, taken from Yang and Swenberg (1991). The paper was dedicated to the memory of J. Neyman, P. S. Puri and E. L. Scott.

Both the Neyman–Puri model and our modification neglect the possibility of a cell's nonlinear repair-misrepair mechanism. Le Cam (1995) pointed out that there is considerable evidence that the repairs are notlinear and some repair is an interaction of two lesions. Solving nonlinear equations in Markov processes is mathematically difficult. There are many problems in this area that need to be studied. Le Cam (1995) wrote, "Neyman was one of the first statisticians to look at applications of statistics in molecular biology."

The preceding examples of Neyman and Brillinger's paper illustrate what Neyman's students wrote in the Foreword in a volume of selected early papers of J. Neyman, edited by students of Neyman (1966), "The interesting feature of the approach used by Neyman is that, in all these papers, the substantive problem is discussed per se and a mathematical model of the phenomenon is constructed. An effort is then made to derive from the structure of the mathematical model new statistical methods particularly adapted to the solution of the problems under consideration. Mere application of standard statistical techniques does not occur in these or later papers."

Neyman was a founding father of modern statistics. Perhaps, the prominence of his fundamental work in the statistics theory overshadows his applied work. In fact, his contribution to and broader impact in applied statistics are equally profound. In Washington, D.C., his applied work is felt through government agencies.

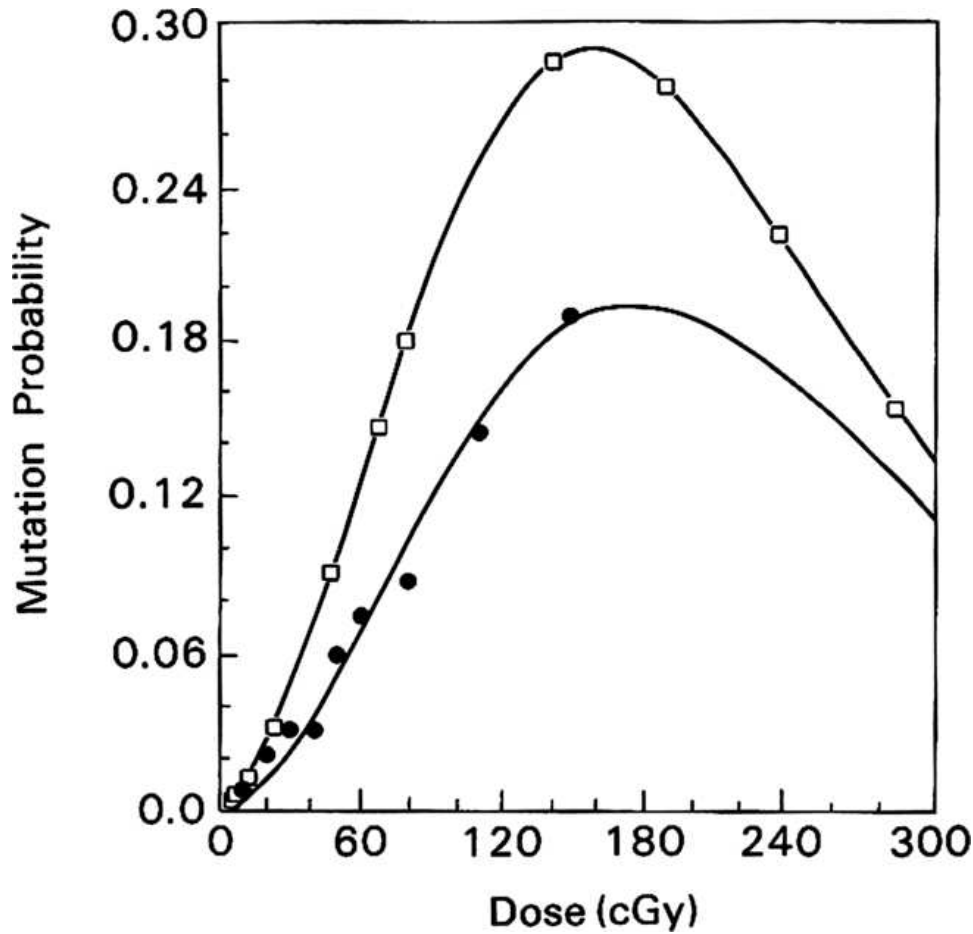

FIG. 2. *Dose-response relationship for pink mutant events per hair after X-irradiation. Filled dot and square denote the experimental mutation fractions. Solid curves denote the least-squares fit of the model mutation probability to the data. [First published as Figure 4 in Yang and Swenberg (1991), "Stochastic modeling of dose-response for single cells in radiation experiments," Math. Scientist, vol. 16, pages 46–65. Copyright © Applied Probability Trust 1991.]*

I end with Neyman's remark on the issue of theoretical and applied statistics:

> This postscript has to deal with the general character of statistical research and with the ties that exist between the pure mathematical theory of statistics and the applied work. I deeply regret the not infrequent emphatic declarations for or against pure theory and for or against work in applications.[7] It is my strong belief that both are important and, certainly, both are interesting. The Berkson–Dantzig–Stein incident just recounted provides an excellent illustration of the view..., The results of Dantzig and Stein* are certainly contributions to pure theory of statistics. Yet, whether the two authors are aware of the fact or not, the theoretical problems they solved originated from difficulties in applied work... (Neyman, 1952, page 268).

[7]Quite recently I was shown some letters regarding myself. One very nice person wrote "I met Neyman. In general he is O.K., but hopelessly mathematical...." The letter of another equally nice person stated: "Once upon a time Neyman did some real work. Now, however, he is interested in applications."
*Refers to Stein's two-stage sequential procedure.

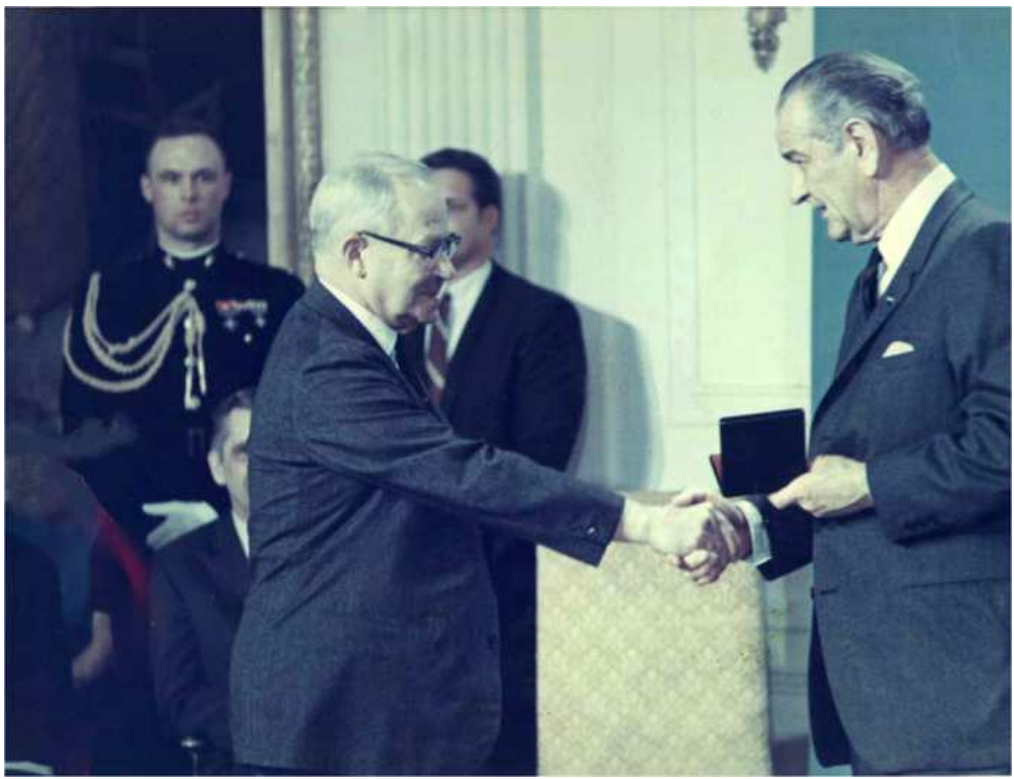

Fig. 3. *J. Neyman, recipient of the National Medal of Science (1968), receiving the medal from President Lyndon Johnson at the White House ceremony on January 17, 1969.*

In Neyman's case, he did both the applied and theoretical work. Neyman's monumental accomplishments did not happen by chance.

(A photo of Neyman receiving the National Medal of Science from President Lyndon Johnson appears on page 75.)

## ACKNOWLEDGMENTS

This material was based on work supported by the National Science Foundation while the author working at the Foundation. Any opinion, finding, and conclusions or recommendations expressed in this material are those of the author and do not necessarily reflect the views of the National Science Foundation.

## REFERENCES

AALEN, O. (1978). Non-parametric inference for a family of counting processes. *Ann. Statist.* **6** 701–726. MR0491547

ALTSHULER, B. (1970). Theory for the measurement of competing risks in animal experiments. *Math. Biosci.* **6** 1–11. MR0266395

BOWLEY, A. L. (1913). Working class households in Reading. *J. Roy. Statist. Soc.* **76** 672–701.

BOWLEY, A. L. (1925). Measurement of the precision attained in sampling. Memorandum, *Bull. Int. Stat. Inst.* **22** 1–62, supplement.

CHIANG, C. L. (1968). *Introduction to Stochastic Processes in Biostatistics*. Wiley, New York.

FIX, E. and NEYMAN, J. (1951). A simple stochastic model of recovery, relapse, dean and loss of patients. *Human Biology* **23** 205–241.

FRANKENBERG-SCHWAGER, M., FRANKENBERG, D., BLOCHER, D. and ADAMCZYK, C. (1980). Repair of DNA double strand breaks in irradiated yeast cells under non growth conditions. *Radiat. Res.* **82** 498–510.

HANSEN, M. (1987). Some history and reminiscences on survey sampling. *Statist. Sci.* **2** 180–190. MR0904033

JENSEN, A. (1925). The report on the representative method in statistics. *Bull. Int. Stat. Inst.* **22** 359–380.

LE CAM, L. (1995). Neyman and stochastic models. *Probab. Math. Statist.* **15** 37–45. MR1369790

LE CAM, L. and NEYMAN, J. (1982). *Probability Models and Cancer*. North-Holland, San Francisco.

MANN, N. R. (1994). In Memoriam: W. Edwards Deming 1900–1993. *J. Amer. Statist. Assoc.* **89** 365–366.

MOORE, C. C. (2007). *Mathematics at Berkeley: A History*. A K Peters Ltd., Wellesley, MA. MR2289685

NEYMAN, J. (1934). On the two different aspects of the representative method: The method of stratified sampling and the method of purposive selection. *J. Roy. Statist. Soc. Ser. B* **97** 558–625.

Neyman, J. (1938). *Lecture Notes and Conferences on Mathematical Statistics*. Graduate School, USDA, Washington, DC.

Neyman, J. (1952). *Lecture Notes and Conferences on Mathematical Statistics and Probability*, 2nd ed. Graduate School, USDA, Washington, DC. MR0052725

Neyman, J. (1960). Indeterminism in science and new demands on statisticians. *J. Amer. Statist. Assoc.* **55** 625–639. MR0116393

Neyman, J. and Students of J.N. at Berkeley (1966). *A Selection of Early Statistical Papers of J. Neyman*. Univ. California Press. MR0222983

Neyman, J. and Puri, P. S. (1976). A structural model radiation effects in living cells. *Proc. Natl. Acad. Sci. USA* **73** 3360–3363. MR0416639

Neyman, J. and Puri, P. S. (1981). A hypothetical stochastic mechanism of radiation effects in single cells. *Proc. Roy. Soc. London B* **213** 139–160.

Tsiates, A. (1975). A nonidentifiabiity aspect of the problem of competing risks. *Proc. Natl. Acad. Sci. USA* **72** 20–22. MR0356425

Yang, G. L. and Swenberg, C. (1991). Stochastic modeling of dose-response for single cells in radiation experiments. *Math. Scientist* **16** 46–65.